\documentclass{article}
\begin{document}

\begin{center}
ANOMALIES FROM IMMERSIONS\\ [.25in] by Juan F. Ospina G.
\end{center}

\begin{center}
ABSTRACT \\ [.25in]
\end{center}

Two forms of anomalies for chiral spinors living on submanifolds of the spacetime are
obtained from the integrality theorem for immersions. The first form of the chiral
anomaly is the usual for chiral spinors living on D-brane and O-plane intersections,
the second form is exotic.

\section{Introduction}
 \setlength{\baselineskip}{20pt}

The anomaly for chiral spinors living on submanifolds of the spacetime may be computed
as the index of an appropiate Dirac operator corresponding to an generalized Spin
complex [1]:
\begin{center}
\setlength{\baselineskip}{40pt}
 { \mathversion{bold} $  index(D)
=(-1)^{\frac{d(d+1)}{2}}\int_{M}ch_{\rho}(V)\frac{ch(S^{+}_{T(M)}-S^{-}_{T(M)})ch(S^{+}_{N(M)}-S^{-}_{N(M)})}{e(T(M))}Td(T(M^{C}))$
}
\end{center}

The computation then produces  [1]:
\begin{center}
\setlength{\baselineskip}{40pt}
 { \mathversion{bold} $  index(D)
=\int_{M}ch_{\rho}(V)\wedge\frac{\hat{A}(R)}{\hat{A}(R')}\wedge{e(R')}$ }
\end{center}

From the other side, the anomaly for chiral spinors living on submanifolds of the
spacetime may be computed as the index of an appropiate Dirac operator twisted with
the superbundle ${\Xi}\rightarrow{\Sigma}$ [2]:

\begin{center}
\setlength{\baselineskip}{40pt}
 { \mathversion{bold} $  index(D)
=(-1)^{\frac{(p+1)(p+2)}{2}}\int_{\Sigma}ch^{+}(\Xi)\wedge\frac{Td(T{\Sigma}{\bigotimes}C))}{\chi(T\Sigma)}$
}
\end{center}
Now the computation then produces [2]:
\begin{center}
\setlength{\baselineskip}{40pt}
 { \mathversion{bold} $  index(D)
=\int_{\Sigma}ch^{+}(E)\wedge{ch^{+}(\overline{E})}\wedge{e^{d(N\Sigma)}}\wedge\frac{\hat{A}(T\Sigma)}{\hat{A}(N\Sigma)}\wedge{\chi(N\Sigma)}$
}
\end{center}

\section{The integrality theorem for immersions}
In this section is presented the following integrality theorem for immersions [3]:

Let X be an n-dimensional closed manifold with a transitive $G_{TX}-structure$, n=2m
even. Let X be immersed into an  (n+k)-dimensional spin manifold Y, such that the
normal bundle $\nu$ carries a $G_{\nu}-structure$. Let $\Phi_{TX}:
X\rightarrow{BG_{TX}}$  and $\Phi_{\nu}: X\rightarrow{BG_{\nu}}$ be the classifying
maps for the tangent and the normal bundle.

Let $\sigma \in {R(\hat{G}_{TX},\hat{H}_{TX})}$ and $V \in {R(\hat{G}_{\nu})}$ such
that (-1,-1) acts trivially on $\sigma\cdot{V}$.  Let  $W \in K^{0}(X)$, then

\begin{center}
\setlength{\baselineskip}{40pt}
 { \mathversion{bold} $
\int_{X}ch(W)\cdot{\Phi_{\nu}^{*}((\pi_{2}^{*})^{-1}ch(V))}\cdot{\Phi_{TX}^{*}(\frac{(\pi_{1}^{*})^{-1}ch(\sigma)}{e|BG_{TX}})}\cdot{\hat{A}(TX)^{2}}=integer$
}
\end{center}
where, $e\in H^{2m}(BSO(2m);Q)$ is the universal Euler class, $ch: R(G)\rightarrow
H^{*}(BG;Q)$ is the universal Chern character, and $\hat{A}(TX)$ is the total
$\hat{A}-class$ of X.

For the proof of the integrality theorem for immersions, the procedure is the
following [3]:

The immersions with certain properties yield structure groups for the manifolds under
consideration, from such structure groups we can to obtain elliptic symbols and then
the corresponding elliptic operators, finally applying the Atiyah-Singer index theorem
for these elliptic operators we can to produce the integrality theorem for immersions.

\section{Anomalies from immersions}

Anomalies are obtained from the integrality theorem for immersions using the following
structure groups for the tangent and normal bundles:

\begin{center}
\setlength{\baselineskip}{40pt}
 { \mathversion{bold} $
G_{TX}=Spin(n)$ }
\end{center}
\begin{center}
\setlength{\baselineskip}{40pt}
 { \mathversion{bold} $
G_{\nu}=Spin^{c}(k)$ }
\end{center}

Then [3]:

\begin{center}
\setlength{\baselineskip}{40pt}
 { \mathversion{bold} $
\Phi_{TX}^{*}(\frac{(\pi_{1}^{*})^{-1}ch(\sigma)}{e|BG_{TX}})\cdot{\hat{A}(TX)^{2}}=\hat{A}(TX)$
}
\end{center}
\begin{center}
\setlength{\baselineskip}{40pt}
 { \mathversion{bold} $
\Phi_{\nu}^{*}((\pi_{2}^{*})^{-1}ch(V))=e^{d(\nu)}\cdot{e(\nu)}\cdot{\hat{A}(\nu)^{-1}}$
}
\end{center}
finally, applying the integrality theorem for immersions is obtained that [3]:
\begin{center}
\setlength{\baselineskip}{40pt}
 { \mathversion{bold} $
\int_{X}ch(W)\cdot{e^{d(\nu)}\cdot{e(\nu)}\cdot{\hat{A}(\nu)^{-1}}}\cdot{\hat{A}(TX)}=integer$
}
\end{center}

This last expression is the usual for the chiral anomaly [1],[2].

An exotic anomaly for chiral spinors that are living on submanifolds of the spacetime
is obtained according to the following procedure:

again,
\begin{center}
\setlength{\baselineskip}{40pt}
 { \mathversion{bold} $
G_{TX}=Spin(n)$ }
\end{center}
\begin{center}
\setlength{\baselineskip}{40pt}
 { \mathversion{bold} $
G_{\nu}=Spin^{c}(k)$ }
\end{center}

but in this case one has the following [3]:
\begin{center}
\setlength{\baselineskip}{40pt}
 { \mathversion{bold} $
\Phi_{TX}^{*}(\frac{(\pi_{1}^{*})^{-1}ch(\sigma)}{e|BG_{TX}})\cdot{\hat{A}(TX)^{2}}=\hat{A}(TX)$
}
\end{center}
\begin{center}
\setlength{\baselineskip}{40pt}
 { \mathversion{bold} $
\Phi_{\nu}^{*}((\pi_{2}^{*})^{-1}ch(V))=2^{l}\cdot{e^{d(\nu)}}\cdot{M(\nu)}$ }
\end{center}
finally applying the integrality theorem for immersions is obtained that:
\begin{center}
\setlength{\baselineskip}{40pt}
 { \mathversion{bold} $
2^{l}\int_{X}ch(W)\cdot{e^{d(\nu)}}\cdot{M(\nu)}\cdot{\hat{A}(TX)}=integer$ }
\end{center}

here $M(\nu)$ is the Mayer class, it is to say, is the multiplicative class for the
power series $\cosh(\frac{x}{2})$, i.e. if we write the Pontrjagin class $p(\nu)$
formally as:
\begin{center}
\setlength{\baselineskip}{40pt}
 { \mathversion{bold} $
p(\nu)=\prod_{j=1}^{l}(1+x_{j}^{2})$ }
\end{center}
then
\begin{center}
\setlength{\baselineskip}{40pt}
 { \mathversion{bold} $
M(\nu)=\prod_{j=1}^{l}\cosh(\frac{x_{j}}{2})$ }
\end{center}

\section{Conclusions}

From the anomaly:
\begin{center}
\setlength{\baselineskip}{40pt}
 { \mathversion{bold} $  index(D)
=\int_{\Sigma}ch^{+}(E)\wedge{ch^{+}(\overline{E})}\wedge{e^{d(N\Sigma)}}\wedge\frac{\hat{A}(T\Sigma)}{\hat{A}(N\Sigma)}\wedge{\chi(N\Sigma)}$
}
\end{center}
the following anomalous RR coupling on the brane-antibrane system is obtained [1],
[2]:
\begin{center}
\setlength{\baselineskip}{40pt}
 { \mathversion{bold} $  Y
=ch^{+}(E)\wedge{e^{\frac{d(N\Sigma)}{2}}}\wedge\sqrt{\frac{\hat{A}(T\Sigma)}{\hat{A}(N\Sigma)}}$
}
\end{center}

Then, the question is,the exotic anomaly is given by:
\begin{center}
\setlength{\baselineskip}{40pt}
 { \mathversion{bold} $
2^{l}\int_{X}ch(W)\cdot{e^{d(\nu)}}\cdot{M(\nu)}\cdot{\hat{A}(TX)}=integer$ }
\end{center}
what is the anomalous coupling that can be obtained from such exotic chiral anomaly?
\section{References}

[1]  C. A. Scrucca  and M. Serone, Nuclear Physics B556 (1999)  hep-th/9903145

[2]  Richard J. Szabo ,  hep-th/0108043

[3]  Christian Bar,  Elliptic Symbols

\end{document}